\documentclass[letterpaper, 10 pt, conference]{ieeeconf}  

\IEEEoverridecommandlockouts                              

\usepackage{graphicx} 
\usepackage{amsmath}
\usepackage{subfigure}
\usepackage{hyperref}
\usepackage[capitalize]{cleveref}

\DeclareMathOperator*{\argmax}{arg\,max}
\usepackage{tikz}
\usepackage[style=ieee, url=false, doi=false, natbib=true, mincitenames=1, maxcitenames=1, maxbibnames=8]{biblatex}
\title{\LARGE \bf
Optimizing Task Completion Time Updates Using POMDPs
}

\author{Duncan Eddy$^{1}$, Esen Yel$^{2}$,
    Emma Passmore$^{1}$,
    Niles Egan$^{1}$,\\
    Grayson Armour$^{1}$,
    Dylan M. Asmar$^{1}$,
    Mykel J. Kochenderfer$^{1}$
\thanks{$^{1}$Duncan Eddy ({\tt\small deddy@stanford.edu}),\newline Emma Passmore ({\tt\small emma@alumni.stanford.edu}),\newline Niles Egan ({\tt\small  negan@stanford.edu}),\newline Grayson Armour ({\tt\small garmour1@alumni.stanford.edu}),\newline Dylan Asmar ({\tt\small asmar@stanford.edu}),\newline Mykel Kochenderfer ({\tt\small mykel@stanford.edu}) are with Stanford University, Stanford, CA, USA}%
\thanks{$^{2}$Esen Yel ({\tt\small yele@rpi.edu}) is with Rensselaer Polytechnic Institute, Troy, NY, USA}%
}

\begin{document}

\maketitle
\thispagestyle{empty}
\pagestyle{empty}

\begin{abstract}

Managing announced task completion times is a fundamental control problem in project management. While extensive research exists on estimating task durations and task scheduling, the problem of when and how to update completion times communicated to stakeholders remains understudied. Organizations must balance announcement accuracy against the costs of frequent timeline updates, which can erode stakeholder trust and trigger costly replanning. Despite the prevalence of this problem, current approaches rely on static predictions or ad-hoc policies that fail to account for the sequential nature of announcement management. In this paper, we formulate the task announcement problem as a Partially Observable Markov Decision Process (POMDP) where the control policy must decide when to update announced completion times based on noisy observations of true task completion. Since most state variables (current time and previous announcements) are fully observable, we leverage the Mixed Observability MDP (MOMDP) framework to enable more efficient policy optimization. Our reward structure captures the dual costs of announcement errors and update frequency, enabling synthesis of optimal announcement control policies. Using off-the-shelf solvers, we generate policies that act as feedback controllers, adaptively managing announcements based on belief state evolution. Simulation results demonstrate significant improvements in both accuracy and announcement stability compared to baseline strategies, achieving up to 75\% reduction in unnecessary updates while maintaining or improving prediction accuracy.



\end{abstract}


\section{Introduction}
\label{intro}
Managing stakeholder announcements of task completion times represents a critical control challenge distinct from the well-studied problem of predicting task durations and task scheduling. Accurate management of task completion announcements is critical for many organizations across various industries, including software engineering~\cite{jones2004software}, construction~\cite{wambeke2011causes}, manufacturing~\cite{Huang2020product}, and aviation~\cite{shumsky1998optimal}. While organizations have access to increasingly sophisticated prediction models, they still face the understudied decision problem of when and how to update communicated timelines as new information becomes available. This announcement control problem requires balancing the competing objectives of maintaining accuracy while minimizing the costs associated with frequent updates.

The distinction between prediction and announcement control is crucial. Even with perfect predictions, organizations must decide whether updating stakeholders is worth the associated costs---customer disappointment, resource reallocation, and erosion of trust. Premature updates based on uncertain early observations may prove unnecessary, while delayed updates can magnify negative impacts. This creates a sequential decision-making problem where the controller must optimally manage the announced completion time based on evolving belief states.

Consider the James Webb Space Telescope. It was initially announced to launch in 2007 at total cost of \$1 billion, but it ultimately launched in 2021 at over \$10 billion~\cite{reichhardt2006us,wang2024jwst}. An independent review found that incremental announcement changes created cascading effects on stakeholder planning and resource allocation, contributing to the cost and schedule creep~\cite{jwstfinalreport2010}. This example illustrates how poor announcement control policies, not just prediction errors, can amplify project costs.

The announcement control problem exhibits several characteristics that make it amenable to control-theoretic approaches~\cite{shumsky1998optimal}. First, the true completion time is partially observable, with observation quality improving over time. Second, each announcement update incurs immediate costs in terms of replanning efforts and reallocation of resources. Third, the cost structure is inherently asymmetric, with different penalties for premature versus delayed updates.

Traditional project management approaches fail to address these dynamics, typically employing static policies (target announced date at all costs) or reactive policies (announce the latest estimate) without considering that delays incurred by replanning make project management a sequential decision problem \cite{usman2014effort}. Single-shot estimates neglect the real costs of timeline volatility on stakeholder trust and operational planning.

To address this gap, we formulate the announcement control problem as a Partially Observable Markov Decision Process (POMDP) \cite{Kaelbing1998planning}. Since most state components (current time and previous announcements) are fully observable, we leverage the Mixed Observability MDP (MOMDP) framework \cite{Ong2009pomdps,ong2010planning} to enable efficient policy synthesis. This formulation captures the inherent uncertainty in task completion while explicitly modeling announcement decisions as control actions with associated costs.

We validate our approach using the QMDP~\cite{littman1995learning} and SARSOP~\cite{Kurniawati2009sarsop} belief-state planning algorithms and compare against heuristic baseline policies. Our framework provides actionable control policies for organizations seeking to optimize stakeholder communication strategies in various industries such as transportation, construction, and software development.

\section{Related Work}

There is clear recognition of the value of accurate task estimation and scheduling from operations research, particularly for resource and cost management. This recognition has led to the compilation of project data across a range of domains, including construction and software engineering, in order to develop models that can improve estimation of total task completion time \citep{Doloi2012,Thiele2021,jauhar2021ms,Flyvbjerg2023}. 

There is a large body of work on estimation of task completion that includes classical methods, such as Critical Path Analysis~\cite{nafkha2016critical} and Earned Value Management~\cite{thiele2025improving}, to more recent machine learning models and algorithms, such as recurrent neural networks (RNNs)~\cite{white2019task}, artificial neural networks (ANNs)~\cite{lishner2022using}, long short-term memory (LSTM)~\cite{Huang2020product}, and hierarchical deep neural networks~\cite{aslan2023hierarchical}. All of these approaches rely on historical training data to predict task completion time, which limits their applicability to novel projects that lack historical completion data for training. It is also worth noting that these methods focus solely on estimating the completion time once at project outset and do not model the effect of updating the announced completion time can have on overall project duration.

In this paper, we focus on cases where the task completion time announcements are made in a sequential manner, instead of a single a one-shot estimate. Recently \citeauthor{kim2025information} studied the problem of optimizing the frequency of wait time updates for customers in an unobservable queue~\cite{kim2025information}. They approach the problem as an analytic optimization of an economic utility function to compute the optimal frequency of updates but it does not account for noise in the wait-time estimates. The sequential, partially-observable nature of the problem, as well as the evidence for the usability of POMDPs in time-based predictions~\cite{maintence_scheduling, Industry-survey}, makes POMDPs a natural candidate for this problem. We use both POMDPs and MOMDPs to frame our problem and use existing solvers to generate project planning policies. All software used to to pose the problem, run experiments, and generate the results in this paper can be found at \url{https://github.com/sisl/POMDP\_Plan}.

\section{Background}

\subsection{Partially Observable Markov Decision Processes}
In a partially observable Markov decision process~\cite{Kaelbing1998planning, Kochenderfer2022}, an agent makes observations $o$ of the true state $s$ and takes control action $a$ at each time step. A POMDP is defined by tuple $\langle \mathcal{S}, \mathcal{A}, \mathcal{O}, T, Z, R, \gamma \rangle$. Here, $\mathcal{S}, \mathcal{A}, \mathcal{O}$ denote the sets of states, control actions, and observations, respectively. The transition function $T(s' \mid s, a)$ models state evolution under control, while the observation function $Z(o \mid s', a)$ captures measurement uncertainty. The reward function $R(s,a)$ encodes control objectives, and $\gamma \in [0,1)$ is the discount factor. The agent maintains belief state $b(s)$ representing the probability of being in state $s$, and the agent seeks a control policy that maximizes expected discounted reward.

\subsection{Mixed Observability Markov Decision Processes}
Mixed observability Markov decision process is a specific form of a POMDP where part of the state space is fully observable, while the rest is hidden~\cite{Ong2009pomdps, ong2010planning}. Therefore, the state $s$ of the problem can be factorized by the fully observable components $x$ and partially observable components $y$: $ s = [x, y] \in \mathcal{X} \times \mathcal{Y}$. A MOMDP is represented as a tuple: $ \langle \mathcal{X}, \mathcal{Y}, \mathcal{A}, \mathcal{O}, T_\mathcal{X}, T_\mathcal{Y}, Z, R, \gamma \rangle$ where $ \mathcal{X} $ is the set of all possible fully observable states, and $\mathcal{Y}$ is the set of all possible partially observable states. Transition function $T_\mathcal{X}$ models how the fully observable states evolve: $ T_\mathcal{X} (x' \mid x, y, a) = P(x' \mid x, y, a)$, and transition function $T_\mathcal{Y}$ models how partially observable states evolve: $ T_\mathcal{X} (y' \mid x, y, a,  x') = P(y' \mid x, y, a,  x') $. The observation function gives the probability of obtaining a certain observation given the state: $ Z(o \mid x', y', a) = P(o \mid x', y', a)  $.

\section{Announcement Control as a POMDP} \label{sec:planning}

The project planning problem is posed as a sequential decision-making problem where the agent (e.g., task planner or project manager) announces the project completion time based on observations that are made as the project progresses. The objective of the agent is to synthesize an announcement control policy that minimizes negative consequences such as financial loss and missed delivery deadlines.

The announcement control problem is fundamentally a Partially Observable Markov Decision Process where the true completion time is hidden from the controller. The formulation has the following components:

\textbf{State and state space:} The state of the problem is characterized by three time values: 1) Current time step $t$, 2) previously announced completion time $T_a^{(t-1)}$, and 3) true completion time $T_s$. Given that the current time step and the previously announced completion time are fully observable to the agent, their values are combined in the observable state: $x(t) = (t, T_a^{t-1})$. On the other hand, the true completion time can never be directly observed; rather, the agent only receives noisy observations of it. Hence, the partially observable state contains the true completion time: $y(t) = T_s$. We assume that projects have a minimum and maximum possible completion times: $T_{\text{min}}$ and $T_{\text{max}}$, respectively. Hence, the observable state space of the problem becomes: $\mathcal{X} = \{ (t, T_a) \mid t \in [0, T_{\text{max}}],\; T_a \in [T_{\text{min}}, T_{\text{max}}] \}$ as the current time can range anywhere between the beginning of the project until the maximum completion time, and the previously announced times can range between $T_{\text{min}}$ and $T_{\text{max}}$. Similarly, the partially observable state space is: $\mathcal{Y} = [T_{\text{min}}, T_{\text{max}}]$ because the true completion time needs to stay between the minimum and maximum completion times.

\textbf{Action and action space:} At each time step $t$, the agent takes a control action by announcing the completion time, so we consider the action as the announced time value: $a(t) = T_a^{t}$. Since this value can range anywhere between $T_{\text{min}}$ and $T_{\text{max}}$, the action space is the same as the partially observable state space with $\mathcal{A} = [T_{\text{min}}, T_{\text{max}}]$. Announcing the same completion time as the previous step models the case of leaving the announcement time unchanged.

\textbf{Observation function and space:} The agent does not have direct access to the true completion time; however, it can receive noisy observations $o_t$ at each time step $t$. We assume that such an observation can be generated by project team members' completion time estimate based on their progress or through task estimation methods. Such observations are expected to be more uncertain earlier in the project timeline, given that more things can change throughout the project course. To capture this effect, our observation model decreases uncertainty as the project approaches completion. Specifically, we model the observation uncertainty as a Gaussian distribution
\begin{equation}
    Z(o_t \mid x', y') = \mathcal{N}(o_t \mid \mu_t, \sigma_t^2)
\end{equation}
where $\mu_t = y'$ and $\sigma_t = (T_s - t)/3$. The standard deviation approaches to zero as the project approaches the completion, reflecting greater certainty closer to the true completion time. Given the limits on the project completion time, we truncate this distribution between $T_{\text{min}}$ and $T_{\text{max}}$ and discretize it. The observation space is the same as the action and partially observable state space with $\mathcal{O} = [T_{\text{min}}, T_{\text{max}}]$.

We selected the Gaussian distribution due to it modeling both underestimation and overestimation of project scope and completion times. For specific application other probability distributions such as the log-normal distribution may be more appropriate~\cite{jorgensen2022}.

\textbf{Transition function:} The transition function $T_{\mathcal{X}}$ defines the probability of advancing to next state $x'$ given the current state and action. Given the observable state is simply the current timestep and the announced completion time, the transition is deterministic
\begin{equation*}
    \begin{aligned}
        & T_{\mathcal{X}} (x' = (t, T_a^t) \mid x, y, a = T_a^t) = 1
        & \text{if } t \geq y \\
        & T_{\mathcal{X}} (x' = (t{+}1, T_a^t) \mid x, y, a = T_a^t) = 1
        & \text{otherwise}
    \end{aligned}
\end{equation*}

The partially observable state transition $T_{\mathcal{Y}}$ models the effect of replanning on the true completion time. It captures the effect that changing the announcement time ($a \neq T_a^{t-1}$) can affect the true completion time due to the disruption cost of replanning. We adopt a simple categorical model of this transition
\begin{equation}
    T_{\mathcal{Y}}(y' \mid x, y, a, x') =
    \begin{cases}
          p_0     & \text{if} y' = T_s \\
          p_s     & \text{if} y' = \min(T_s + \delta_s,\; T_{\max}) \\
          p_\ell  & \text{if} y' = \min(T_s + \delta_\ell,\; T_{\max}) \\
          0       & \text{otherwise}
    \end{cases}
    \label{eq:transition_y}
\end{equation}
where $p_0 + p_s + p_\ell = 1$, $\delta_s < \delta_\ell$. The parameter $p_0$ is the probability that replanning has no effect on the true completion time, while $p_s$ and $p_\ell$ govern the likelihood of small $\delta_s$ and large $\delta_\ell$ delays, respectively. This stochastic transition captures the intuition that revising a project plan mid-execution can introduce delay due to resource reallocation, team disruption, or rework. This simple model could be replaced with domain-specific models or learned delay models.

Finally, when the agent retains its previous announcement ($a = T_a^{t-1}$), makes its initial announcement ($t = 0$), or the project has completed ($t \geq y$), the true completion time remains unchanged
\begin{equation*}
    T_{\mathcal{Y}}(y' = y \mid x, y, a, x') = 1\; \text{if } a = T_a^{t-1},\; t = 0, \text{or } t \geq y
\end{equation*}

\textbf{Reward function:} The reward function allows us to encode the control objectives of the agent. The goal of the agent is to make announcements that are close to the true completion time (i.e., minimizing the estimation error), and to minimize the number of completion time estimate updates. To encode these objectives, we use the following reward function
\begin{align*}
    R(x, y, a) = \begin{cases}
        0 & \text{if} t = T_{\text{max}}-1 \\
        0 & \text{if} t \geq T_s  \\
        -\lambda_e |a - y| - \lambda_c \textbf{1}_c - \lambda_f\textbf{1}_f & \text{otherwise}
    \end{cases}
\end{align*}
where $\textbf{1}_c$ is an indicator function for detecting whether or not the agent is changing the announcement time to something other than the true completion time
\begin{align*}
     \textbf{1}_c = \begin{cases}
         1 & \text{if} a \neq T_a^{t-1} \text{and} a \neq y \\
         0 & \text{otherwise}
     \end{cases}
\end{align*}
and $\textbf{1}_f$ is an indicator function to detect if the announcement is the true completion time when the project is finished
\begin{align*}
     \textbf{1}_f = \begin{cases}
         1 & \text{if} a \neq y \text{and} t = y\\
         0 & \text{otherwise}
     \end{cases}
\end{align*}

The term $-\lambda_e | a - y |$ penalizes announcements that are different than the true completion time, the term $-\lambda_c \textbf{1}_c$ induces additional penalty for changing the announcement time unless to announce the correct completion time, and the term $-\lambda_f\textbf{1}_f$ induces additional penalty for not announcing the true completion time once the task or the project is completed. The coefficients $\lambda_e$, $\lambda_c$, and $\lambda_f$ allow us to balance these components of the reward function.

This MOMDP formulation of the underlying POMDP enables the use of specialized solvers that can take advantage of the mixed observability structure for improved efficiency, while maintaining the ability to synthesize optimal announcement control policies that balance accuracy and stability.

\section{Results}

Our experiments involve project planning problems of four sizes: small ($T_\text{min} = 2$ and $T_\text{max} = 13$), medium ($T_\text{min} = 2$ and $T_\text{max} = 26$), large ($T_\text{min} = 2$ and $T_\text{max} = 39$), and extra large ($T_\text{min} = 2$ and $T_\text{max} = 52$). These problems have state space sizes of $\mid\mathcal{S}\mid = 1,862$, 16,520, 56,316, and 135,252, respectively. We chose these values to resemble projects that are quarter, half-year, three quarters, and a year long with weekly completion time announcements. Based on the project type, the same problem set up can be used for projects of different durations.

The announcement control problem formulation requires the user to make some parameter choices. We use a discount factor $\gamma=0.98$ to prioritize long-term benefits of control actions taken over time steps. In the reward function, we use $\lambda_e = 8$, $\lambda_c = 2$, and $\lambda_f = 1000$ to balance announcement accuracy and frequency. Small delays $\delta_s = 1$ occur with $p_s = 0.4$ and large delays $\delta_\ell = 3$ with $p_\ell = 0.1$, with no delay incurred with probability $p_0 = 0.5$. These are notional parameter values selected to demonstrate the formulation; the underlying problem structure and solution approach remain valid for more complex parameterizations or domain-specific models of delay and observation uncertainty.

\subsection{Solvers}
To generate announcement control policies, we use two off-the-shelf solvers that leverage different aspects of our problem structure. Since the announcement control problem is fundamentally a POMDP with mixed observability, we apply QMDP~\cite{littman1995learning} directly to the POMDP formulation and SARSOP~\cite{Kurniawati2009sarsop} to the MOMDP formulation to exploit the efficiency gains from the factored state representation.

\textbf{QMDP:}
QMDP~\cite{littman1995learning} is an offline solver that iteratively updates a set of alpha vectors that defines a value function and a control policy. In QMDP, the alpha vectors are updated assuming that the system will have full observability after the first step. Due to this assumption, QMDP results in an upper bound to the true value function after a finite number of steps~\cite{Hauskrecht2000value, Kochenderfer2022}.

\textbf{SARSOP:} SARSOP~\cite{Kurniawati2009sarsop} is an offline solver that improves the computational efficiency of point-based value iteration methods by focusing on determining the optimally reachable belief space to be used when performing belief updates. Given that the optimal control policy is unknown to the agent, the optimally reachable belief space is also unknown. SARSOP approaches this problem by iteratively approximating the optimally reachable belief space. When applied to our MOMDP formulation, SARSOP can efficiently exploit the mixed observability structure to synthesize high-quality announcement control policies.

\subsection{Baselines}
We compare the control policies generated using the POMDP formulation to two baseline announcement strategies:

\textbf{Last observed:} This deterministic policy makes project completion time announcements solely based on observations. Namely, at every time step the project completion time is announced to be the last observation: $a(t) = o_{t-1}$.

\textbf{Most likely:} This policy keeps a belief over the true completion time and announces the most likely completion time at every time step: $ a(t) = \bar{T}_s $ where $\bar{s} = (\bar{t}, \bar{T}_a^{t-1}, \bar{T}_s) = \argmax_{s} b(s) $ is the most likely state.


\subsection{Simulation Results}
To evaluate and compare these announcement control policies, we ran $1000$ simulations for each problem size and policy. While we would have preferred to use real project data, we were unable to find public datasets that included regular updates over multiple episodes with both estimated and true completion times; available datasets contained only initial and final estimates or sparse updates from large-scale projects. For fair comparison, each of our simulation runs for different policies use the same sequence of observations, generated using the observation function given in~\cref{sec:planning}. All simulations used an AMD EPYC 9554 64-Core workstation with 512 GB of RAM running Ubuntu 24.04 LTS and Julia 1.11.6. This computer had significantly more memory than what is needed for solving the problem as the only memory requirements are for storing the policy and performing belief updates.

\begin{figure}[t]
    \centering
    \includegraphics[width=0.999\linewidth]{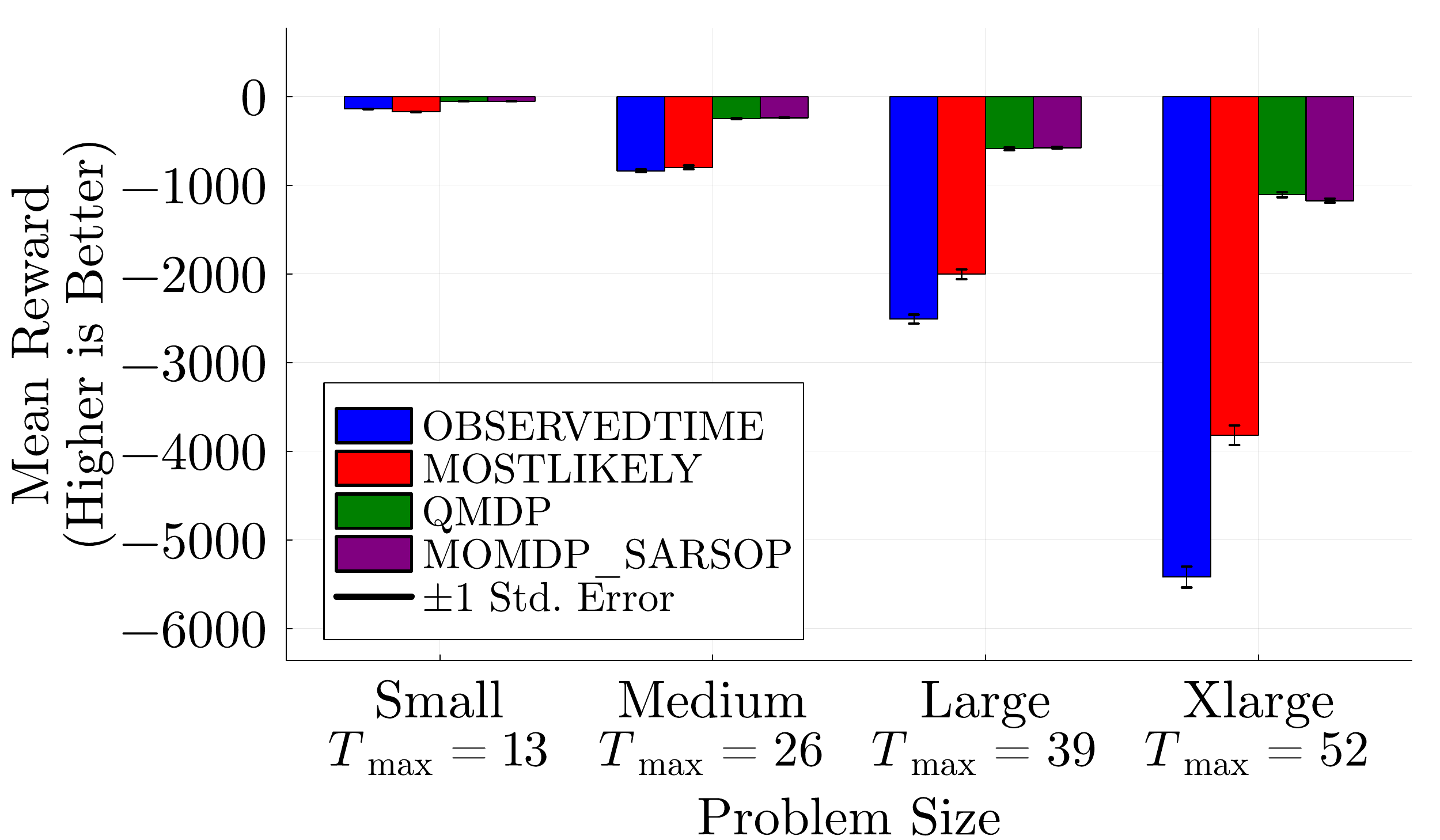}
    \caption{Comparison of reward values with different policies.}
    \label{fig:reward-comparison}
\end{figure}

\Cref{fig:reward-comparison} compares the mean and standard deviation of the reward value over $1000$ simulations with \textsc{qmdp} and \textsc{sarsop} control policies, and with the baseline policies (\textsc{mostlikely} and \textsc{observedtime}). In all problem sizes, \textsc{qmdp} and \textsc{sarsop} perform comparably when it comes to average reward, and both outperform the baseline policies. We also observe that increasing the problem size causes lower average rewards for all policies. This is expected given the increased action and state space sizes, and initial observations are more likely to be erroneous in larger problems.

\begin{figure}
    \centering
    \includegraphics[width=0.95\linewidth]{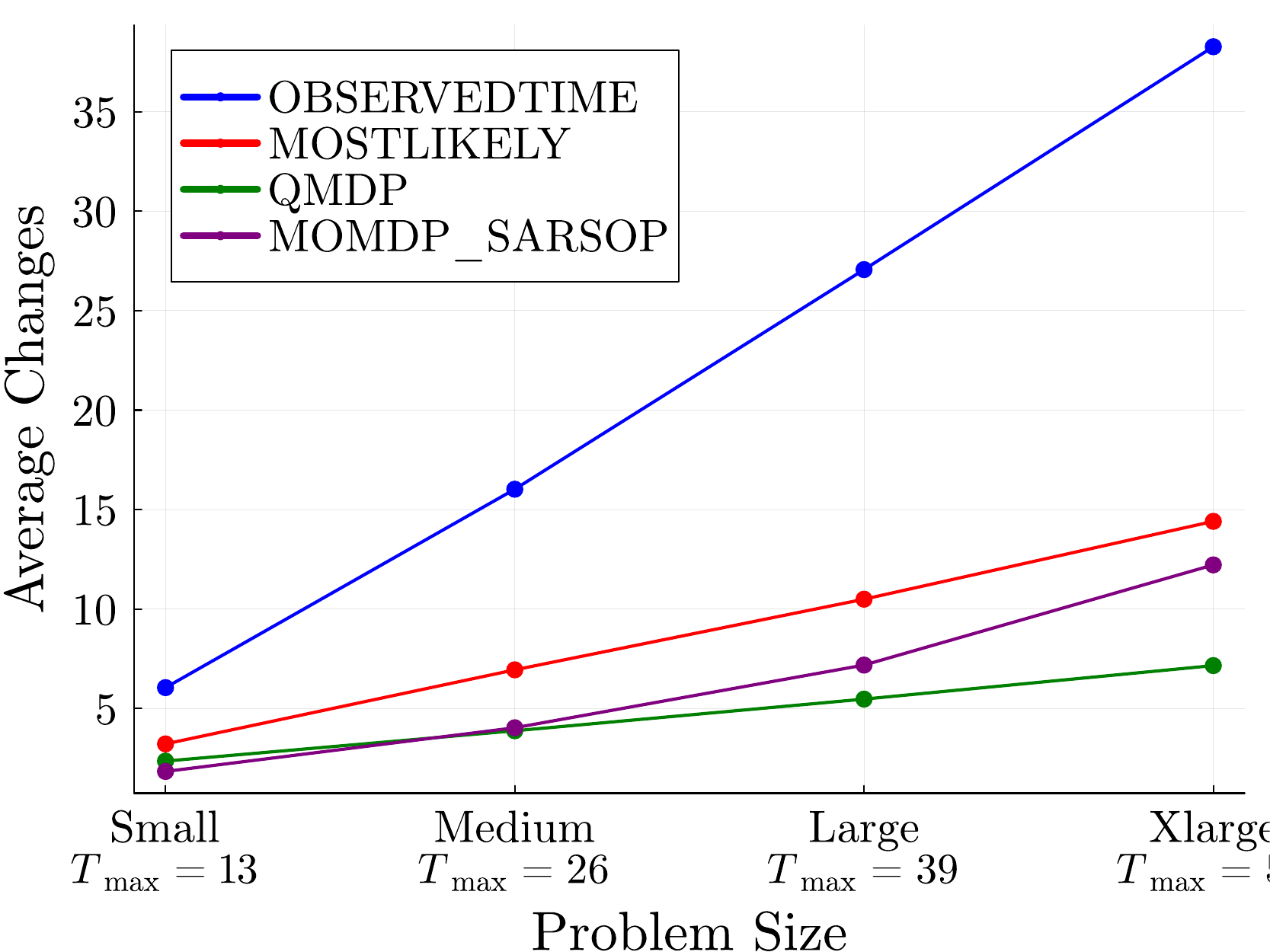}
    \caption{Comparison of average number of announcement changes with different policies.}
    \label{fig:announcement_change}
\end{figure}

Another metric we use to compare the announcement control policies is the average number of changes in the announced completion time. A good control policy should estimate the true completion time accurately and should not change the announcement frequently because such changes can cause replanning of resource allocations, reduce productivity, and incur other change costs. \Cref{fig:announcement_change} shows that in all problem sizes, \textsc{qmdp} and \textsc{sarsop} result in fewer average number of announcement changes than the baseline policies. This result is expected as the announcement changes are penalized in the reward function. For the \textsc{observedtime} policy, it is expected to have frequent announcement changes because it announces the observed completion times directly, resulting in an announcement change whenever a new observation comes in. The \textsc{mostlikely} policy also results in more frequent announcement changes as the announcement is updated whenever the most likely state changes during belief updates. \textsc{qmdp} and \textsc{sarsop} perform similarly in small problems, and \textsc{qmdp} outperforms \textsc{sarsop} in larger problems.

\begin{figure}[t]
    \centering
    \includegraphics[width=0.995\linewidth]{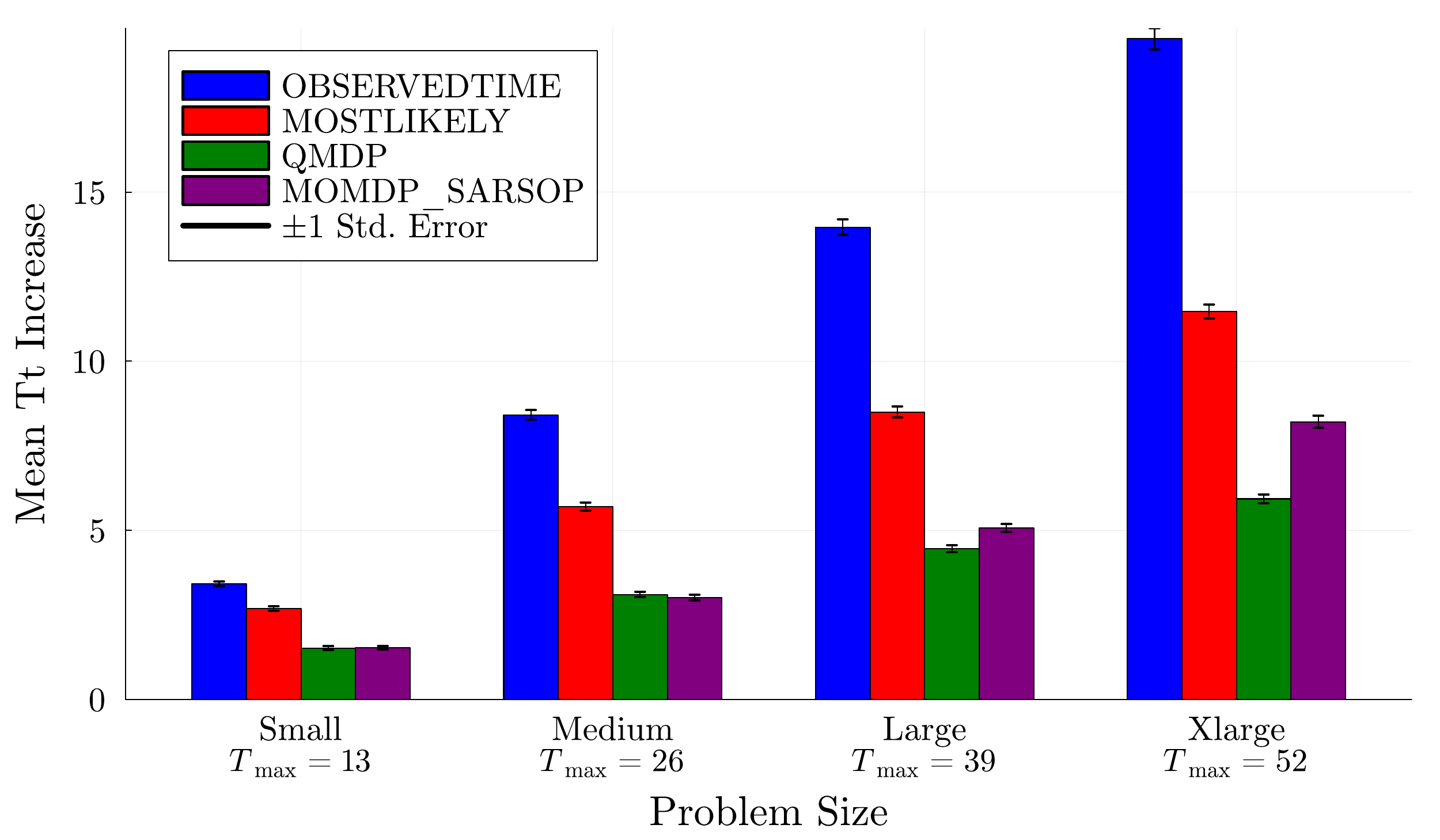}
    \caption{Comparison of average increase in project completion time due to controller announcement time changes}
    \label{fig:tt-increase}
\end{figure}

\begin{figure}[t]
    \centering
    \includegraphics[width=0.95\linewidth]{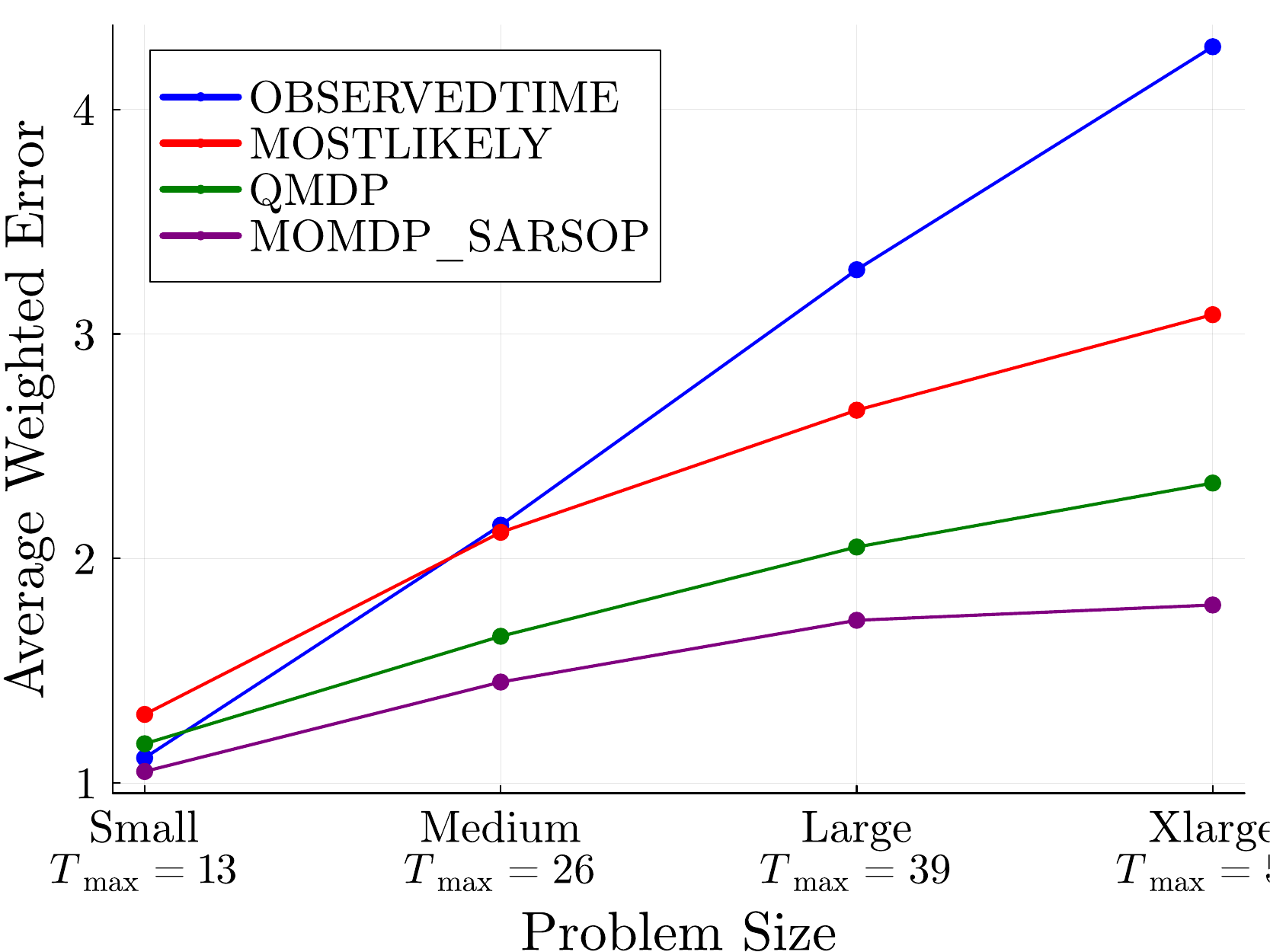}
    \caption{Comparison of average error over the planning horizon with different policies. The value is computed as the average over all simulations of the sum total error across all planning steps.}
    \label{fig:avg-error}
\end{figure}

\Cref{fig:tt-increase} shows the average amount the true completion time increases due the announcement time changes of different controllers for different problem sizes. \textsc{observedtime} and \textsc{mostlikely} frequently change the announced completion time as shown in \Cref{fig:announcement_change} which incurs added costs to replanning, delaying the project overall. \textsc{qmdp} and \textsc{sarsop} complete projects with fewer delays due to actively accounting for the possibility of replanning-induced delays. As confirmed by \Cref{fig:avg-error} the \textsc{qmdp} and \textsc{sarsop} controllers accept that the announced completion time may be incorrect for longer periods of time until it makes sense to change the announced completion time to reduce the number of changed-induced delays.



\subsection{Reward Parameter Tuning: Pareto Analysis of $\lambda_e$ and $\lambda_c$}

The reward function parameters $\lambda_e$ and $\lambda_c$ represent a trade-off in announcement control: the balance between prediction accuracy given evolving information and announcement stability. Increasing $\lambda_e$ further penalizes inaccurate announcements, while increasing $\lambda_c$ penalizes frequent changes to announced deadlines. To systematically explore this trade-off space and identify optimal parameter configurations for the control policy, we performed parameter sweeps over both $\lambda_e$ and $\lambda_c$ using the values $[0.5, 1.0, 2.0, 3.0, 5.0, 8.0, 12.0, 20.0]$, resulting in 64 unique parameter combinations. For each combination, we solved a \textsc{qmdp} control policy using the medium-sized problem configuration ($T_{\text{min}} = 2$, $T_{\text{max}} = 26$) and evaluated it over 100 simulation runs using consistent replay data to ensure fair comparison. For all trials, $\lambda_f$ was held constant at 1000.

\begin{figure}[t]
    \centering
    \includegraphics[width=0.95\linewidth]{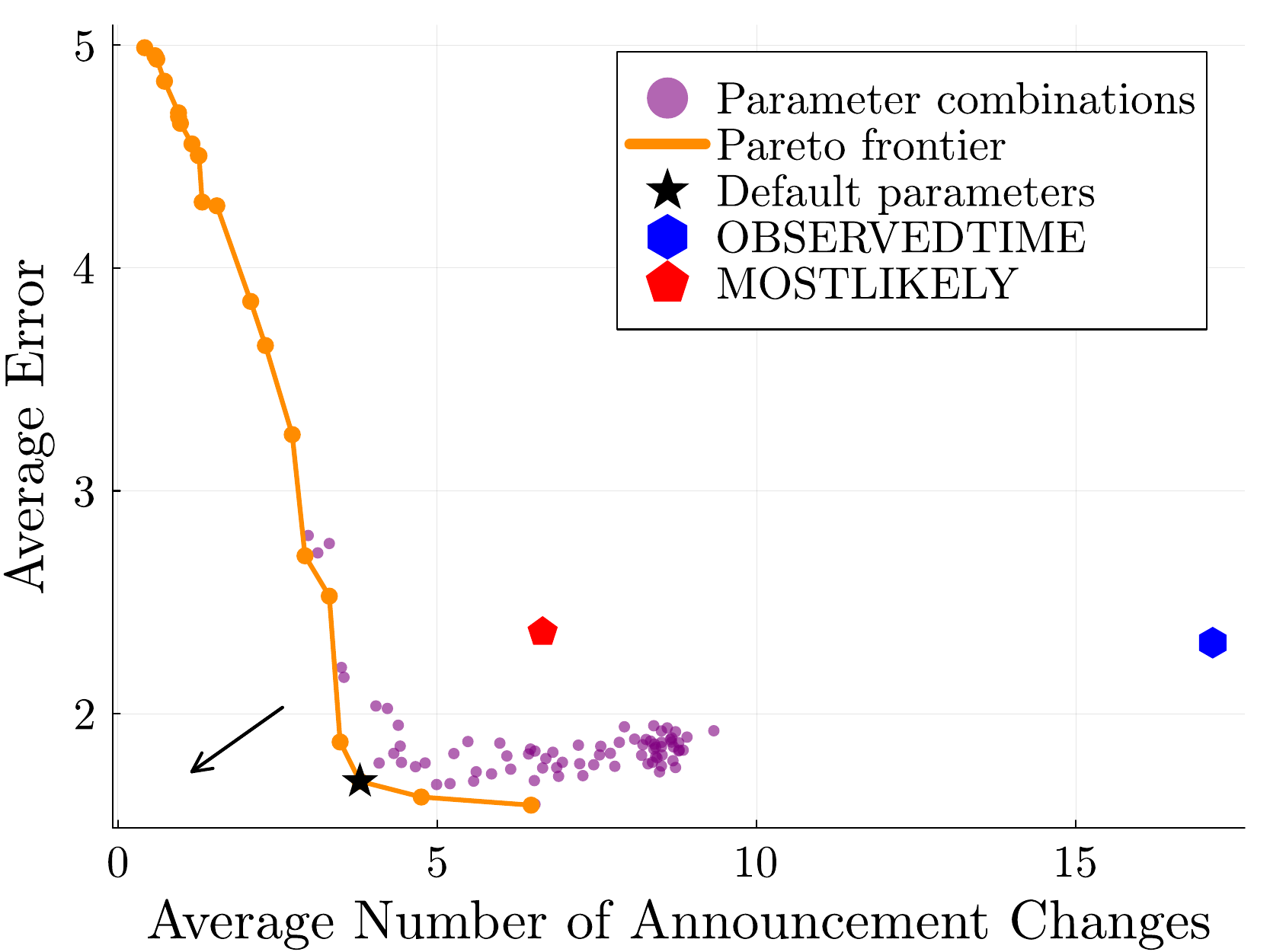}
    \caption{Pareto frontier analysis showing the trade-off between average prediction error and number of announcement changes for \textsc{qmdp} policies. Each point represents a different $(\lambda_e, \lambda_c)$ parameter combination. The gold line indicates the Pareto-optimal configurations, while baseline policies are shown for reference. The black arrow provides the direction of global improvement.}
    \label{fig:pareto-frontier}
\end{figure}

\Cref{fig:pareto-frontier} presents the hyperparameter sweep analysis. This figure shows a clear trade-off between prediction accuracy and announcement frequency in the control policy, with the Pareto frontier demonstrating that achieving very low error rates requires accepting more frequent updates. Our default parameter choice ($\lambda_e = 8$, $\lambda_c = 2$, highlighted with a black star) balances between a low error rate and fewer changes. \textsc{observedtime} and \textsc{mostlikely} perform worse than any parameter combination on the frontier, confirming the value of the optimized announcement control policies over estimation-only approaches.

\subsection{Belief Evolution Analysis}
We analyze the evolution of beliefs for the simulated software engineering project. We consider the scenario of a complex software engineering project, that could take up to 1 year (52 weeks) to complete. The initial true completion time is 22 weeks. The project managers run an agile process and meet weekly with the engineers to get a noisy estimate of the project completion time. Then based on different control policies make a decision to update project stakeholders about the true project completion time and realign engineers around the new completion estimate.

\begin{figure*}[!th]
    \centering
    
    \subfigure[\textsc{observedtime}]{
        \includegraphics[width=0.475\textwidth]{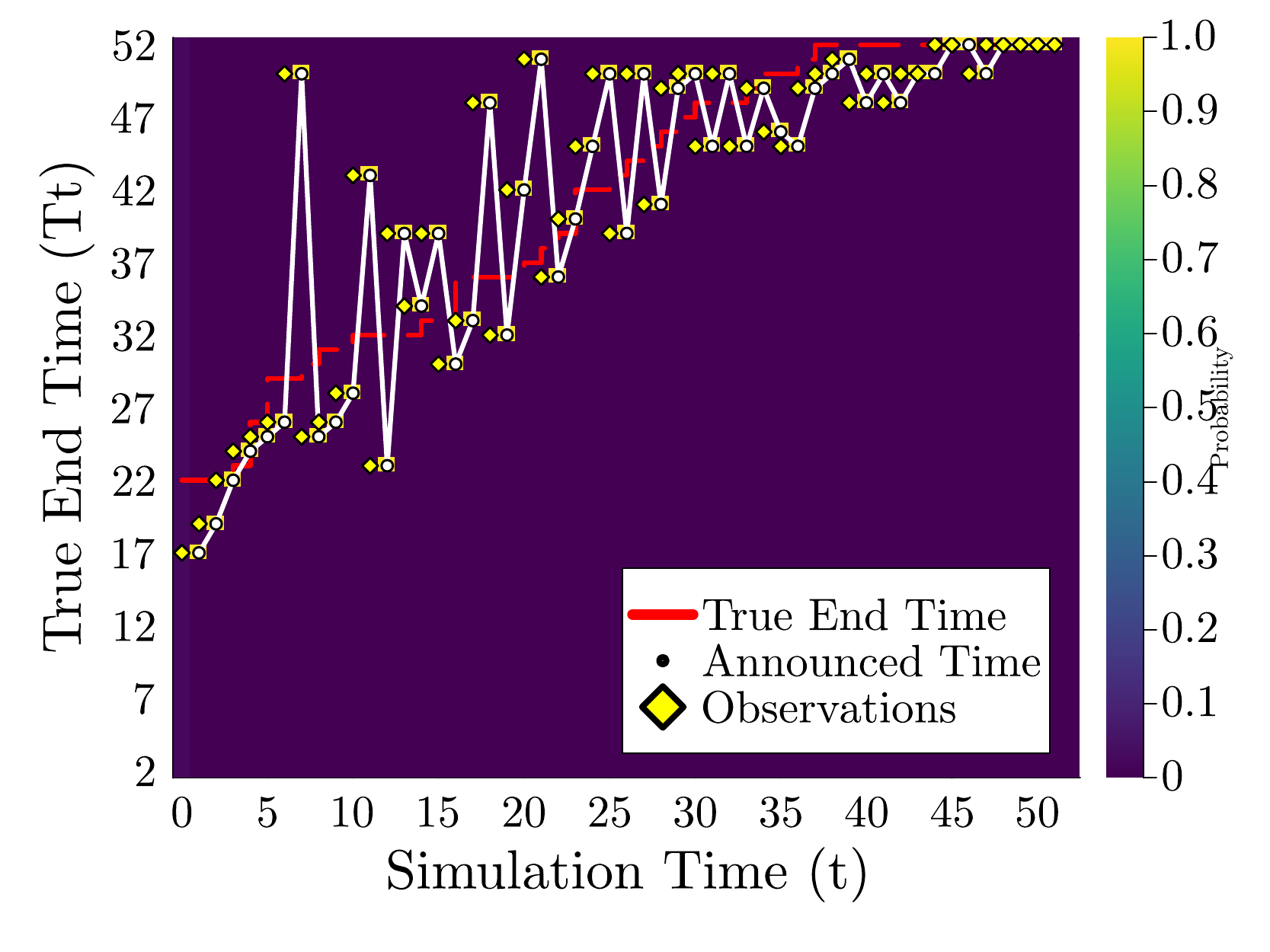}
        \label{fig:belief-xl-observed}
    }
    \subfigure[\textsc{mostlikely}]{
        \includegraphics[width=0.475\textwidth]{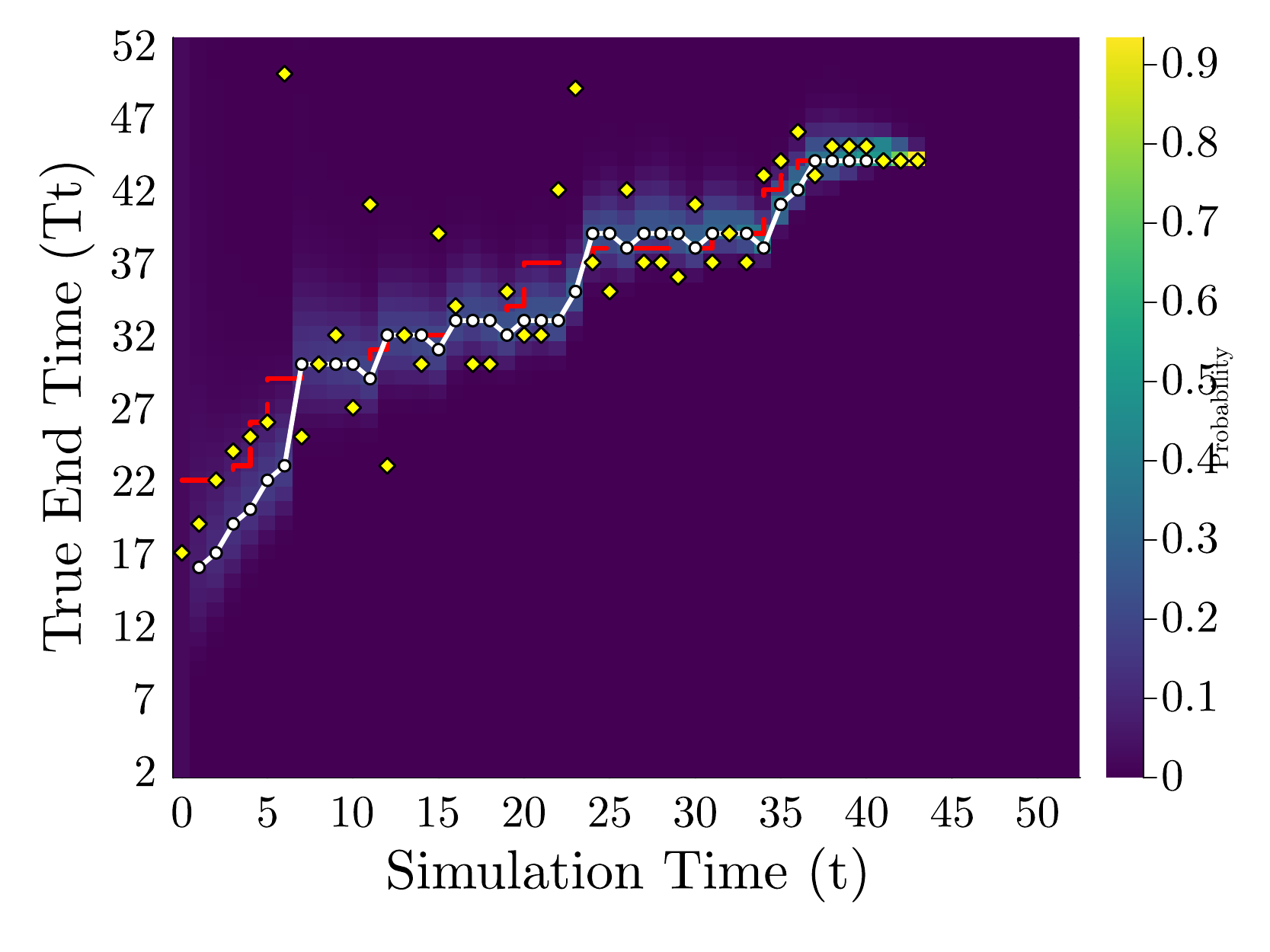} 
        \label{fig:belief-xl-mostlikely}
    } 
    \subfigure[\textsc{qmdp}]{
        \includegraphics[width=0.475\textwidth]{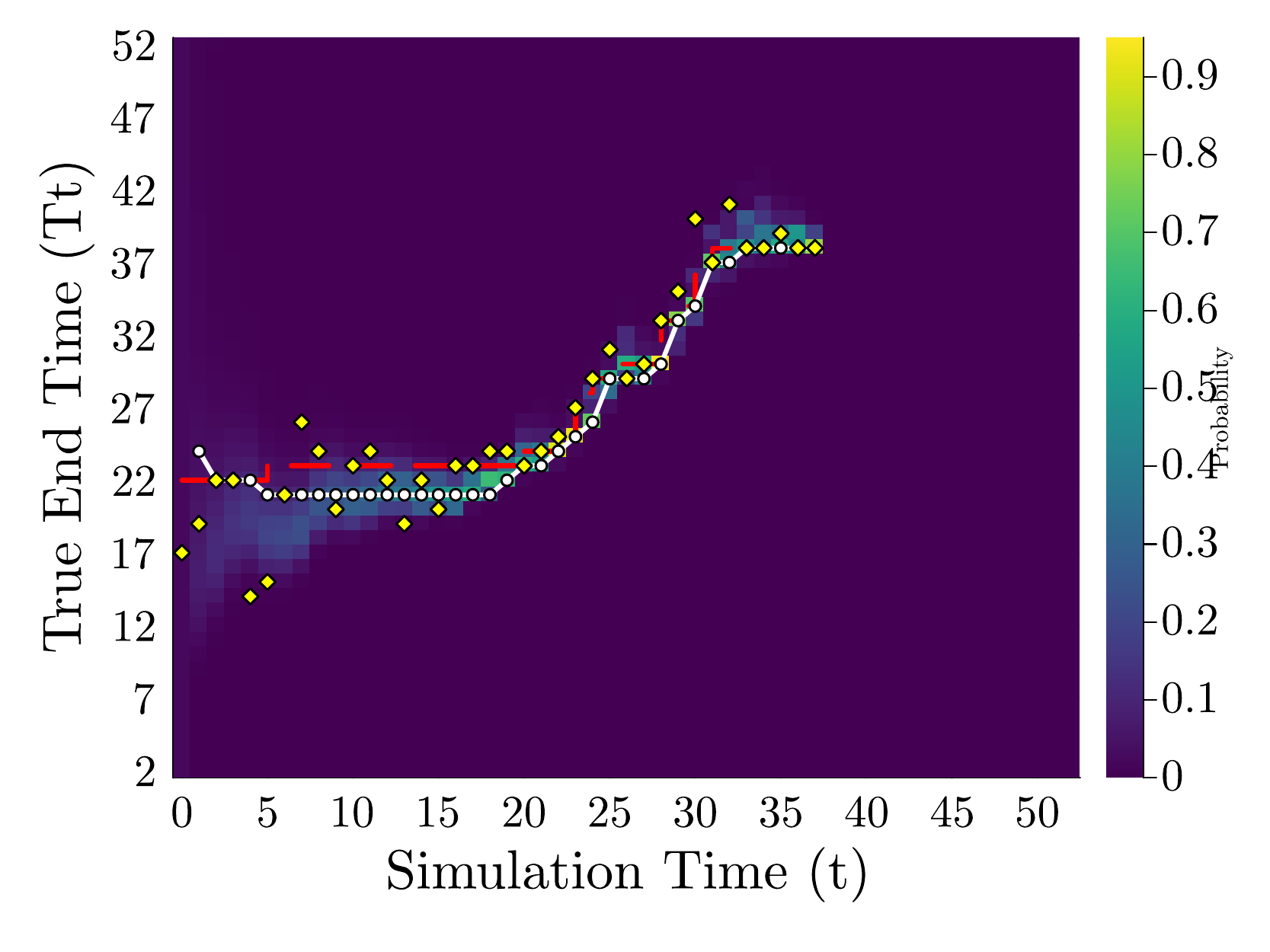}
        \label{fig:belief-xl-qmdp}
    } 
    \subfigure[\textsc{sarsop}]{
        \includegraphics[width=0.475\textwidth]{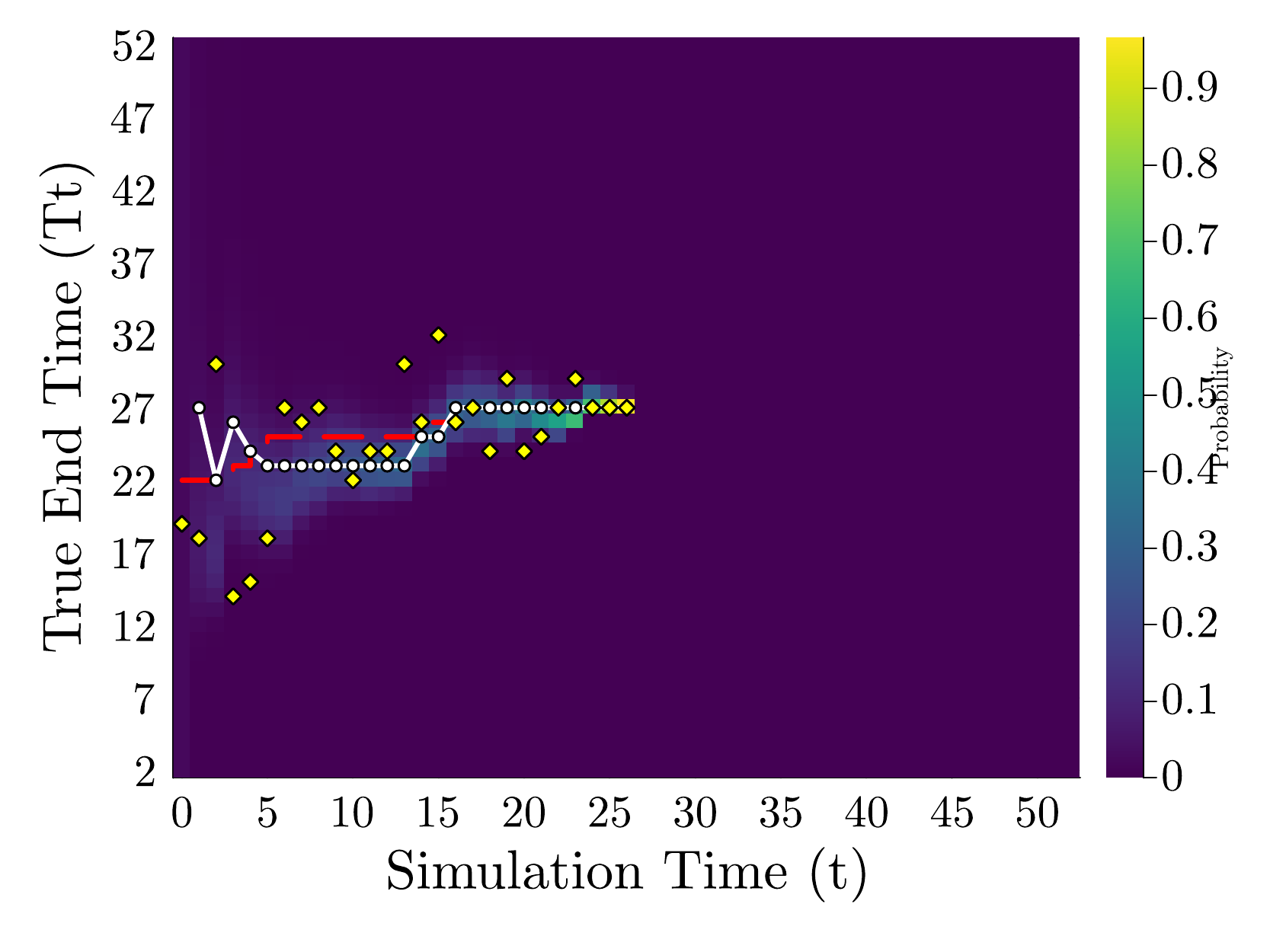} 
        \label{fig:belief-xl-sarsop}
    } 
    \caption{Belief evolution with observed completion times and announcements for a year-long software engineering project with initial true completion estimate of $T_s = 22$. The planning based methods (bottom row) display fewer changes in the announced project completion time over the planning horizon significantly reducing increase in project completion time due to incurred replanning cost.}
    \label{fig:belief-evol-xl}
\end{figure*}

\Cref{fig:belief-evol-xl} shows results of executing this project with four different controllers. Because of the larger problem size, the initial completion estimates have large uncertainty at the beginning of the project, increasing the difficulty of the control problem considerably. We see that for both the \textsc{observedtime} and \textsc{mostlikely} policies, they update their project completion estimates often in reaction to new understandings of the true completion time. However, this replanning does sometimes incur a cost due to lost-engineer time being involved in rescoping and project-alignment meetings. We see that these replanning costs increase the true completion time to 52 weeks for the the for the \textsc{observedtime} policy and 36 weeks for \textsc{mostlikely} controller. Increasing the project duration 136\% and 63\%, respectively.

On the other hand we see that the \textsc{qmdp} and \textsc{sarsop} controllers, hold their initial announcements for nearly the entire project duration, with \textsc{sarsop} changing the initial estimate once at the start, and only at the very last time step before project completion.

\section{Conclusion and Future Work}

In this paper, we introduce a POMDP formulation for the understudied problem of announcement control in task completion forecasting. While extensive research exists on predicting task durations, the distinct decision-making problem of when and how to update stakeholder-communicated timelines has received limited attention. We address this gap by formulating the announcement control problem as a POMDP where the true completion time is partially observable. Since most state components (current time and previous announcements) are fully observable, we leverage the MOMDP framework for efficient policy synthesis.

We use SARSOP and QMDP as policy solvers and compare the synthesized announcement control policies against two baseline strategies. Our simulation results reveal that POMDP-based announcement control has significant potential to improve project planning practices. The control policies achieve superior performance by balancing announcement accuracy against update frequency, addressing the fundamental trade-off in stakeholder communication.

Future work can extend this framework with online planners such as POMCP~\cite{Silver2010montecarlo}, incorporate more complex dynamics, measurement, and reward model structures, extend to more complex project scenarios with interdependent tasks, and deploy the announcement control framework on real project management data to assess operational performance.

\renewcommand*{\bibfont}{\normalfont\footnotesize}
\printbibliography

@book{Kochenderfer2022,
   author = {Mykel J. Kochenderfer and Tim A. Wheeler and Kyle H. Wray},
   city = {Cambridge},
   publisher = {MIT Press},
   title = {{Algorithms for Decision Making}},
   year = {2022},
}

@article{Kaelbing1998planning,
author = {Leslie Pack Kaelbling and Michael L. Littman and Anthony R. Cassandra},
title = {{Planning and Acting in Partially Observable Stochastic Domains}},
journal = {Artificial Intelligence},
volume = {101},
number = {1},
pages = {99-134},
year = {1998}
}

@INPROCEEDINGS{Ong2009pomdps,
    AUTHOR    = {S. C. W. Ong AND S. W. Png AND D. Hsu AND W. S. Lee},
    TITLE     = {{{POMDP}s for Robotic Tasks with Mixed Observability}},
    BOOKTITLE = {Robotics: Science and Systems},
    YEAR      = {2009},
    MONTH     = {June},
    DOI       = {10.15607/RSS.2009.V.026} 
}

@article{ong2010planning,
  title={{Planning Under Uncertainty for Robotic Tasks with Mixed Observability}},
  author={Ong, Sylvie CW and Png, Shao Wei and Hsu, David and Lee, Wee Sun},
  journal={The International Journal of Robotics Research},
  volume={29},
  number={8},
  pages={1053--1068},
  year={2010},
  publisher={SAGE Publications Sage UK: London, England}
}

@incollection{littman1995learning,
  title={{Learning Policies for Partially Observable Environments: Scaling Up}},
  author={Littman, Michael L and Cassandra, Anthony R and Kaelbling, Leslie Pack},
  booktitle={Machine Learning},
  pages={362--370},
  year={1995},
  publisher={Elsevier}
}

@inproceedings{Kurniawati2009sarsop,
  author={Kurniawati, Hanna and Hsu, David and Lee, Wee Sun},
  booktitle={{Robotics: Science and Systems}}, 
  title={{SARSOP: Efficient Point-Based POMDP Planning by Approximating Optimally Reachable Belief Spaces}}, 
  year={2009},
  volume={},
  number={},
  pages={65-72},
  doi={}}

@article{Hauskrecht2000value,
author = {Hauskrecht, Milos},
title = {{Value-Function Approximations for Partially Observable Markov Decision Processes}},
year = {2000},
issue_date = {August 2000},
publisher = {AI Access Foundation},
volume = {13},
number = {1},
journal = {Journal of Artificial Intelligence Research},
month = aug,
pages = {33–94},
numpages = {62}
}

@inproceedings{Silver2010montecarlo,
 author = {Silver, David and Veness, Joel},
 booktitle = {Advances in Neural Information Processing Systems},
 title = {{Monte-Carlo Planning in Large POMDPs}},
 year = {2010}
}

@inproceedings{white2019task,
  title={{Task Duration Estimation}},
  author={White, Ryen W and Hassan Awadallah, Ahmed},
  booktitle={ACM International Conference on Web Search and Data Mining},
  pages={636--644},
  year={2019}
}

@article{lishner2022using,
  title={{Using an Artificial Neural Network for Improving the Prediction of Project Duration}},
  author={Lishner, Itai and Shtub, Avraham},
  journal={Mathematics},
  volume={10},
  number={22},
  pages={4189},
  year={2022},
  publisher={MDPI}
}

@article{Huang2020product,
title = {{Product Completion Time Prediction Using A Hybrid Approach Combining Deep Learning and System Model}},
journal = {Journal of Manufacturing Systems},
volume = {57},
pages = {311-322},
year = {2020},
doi = {https://doi.org/10.1016/j.jmsy.2020.10.006},
author = {Jing Huang and Qing Chang and Jorge Arinez},
}

@article{aslan2023hierarchical,
  title={{Hierarchical Ensemble Deep Learning for Data-Driven Lead Time Prediction}},
  author={Aslan, Ayse and Vasantha, Gokula and El-Raoui, Hanane and Quigley, John and Hanson, Jack and Corney, Jonathan and Sherlock, Andrew},
  journal={The International Journal of Advanced Manufacturing Technology},
  volume={128},
  number={9},
  pages={4169--4188},
  year={2023},
  publisher={Springer}
}

@inproceedings{jauhar2021ms,
  title={{MS-LaTTE: A Dataset of Where and When To-Do Tasks are Completed}},
  author={Jauhar, Sujay Kumar and Chandrasekaran, Nirupama and Gamon, Michael and White, Ryen W},
  booktitle={Conference on Language Resources and Evaluation},
  year={2022}
}

@inproceedings{SARSOP,
  author={Kurniawati, Hanna and Hsu, David and Lee, Wee Sun},
  booktitle={Robotics: Science and Systems}, 
  title={{SARSOP: Efficient Point-Based POMDP Planning by Approximating Optimally Reachable Belief Spaces}}, 
  year={2009}}

@article{maintence_scheduling,
title = {{A POMDP Framework for Integrated Scheduling of Infrastructure Maintenance and Inspection}},
journal = {Computers \& Chemical Engineering},
volume = {112},
pages = {239-252},
year = {2018},
doi = {https://doi.org/10.1016/j.compchemeng.2018.02.015},
url = {https://www.sciencedirect.com/science/article/pii/S0098135418300826},
author = {Jong Woo Kim and Go Bong Choi and Jong Min Lee}
}

@article{Doloi2012,
  author = {Doloi, H. and Sawhney, A. and Iyer, K. C. and Rentala, S.},
  title = {{Analysing Factors Affecting Delays in Indian Construction Projects}},
  journal = {International Journal of Project Management},
  volume = {30},
  number = {4},
  pages = {479-489},
  year = {2012},
  doi = {10.1016/j.ijproman.2011.10.004}
}

@book{Flyvbjerg2023,
  author = {Flyvbjerg, B. and Gardner, D.},
  title = {{How Big Things Get Done: The Surprising Factors that Determine the Fate of Every Project, from Home Renovations to Space Exploration and Everything in Between}},
  publisher = {Crown Currency},
  year = {2023},
  address = {New York}
}

@article{Thiele2021,
  author = {Thiele, B. and Ryan, M. and Abbasi, A.},
  title = {{Developing a Dataset of Real Projects for Portfolio, Program and Project Control Management Research}},
  journal = {Data in Brief},
  volume = {34},
  pages = {106659},
  year = {2021},
  doi = {10.1016/j.dib.2020.106659}
}

@article{Industry-survey,
title = {{Modern Machine Learning Tools for Monitoring and Control of Industrial Processes: A Survey}},
journal = {IFAC-PapersOnLine},
volume = {53},
number = {2},
pages = {218-229},
year = {2020},
doi = {https://doi.org/10.1016/j.ifacol.2020.12.126},
author = {R. Bhushan Gopaluni and Aditya Tulsyan and Benoit Chachuat and Biao Huang and Jong Min Lee and Faraz Amjad and Seshu Kumar Damarla and Jong Woo Kim and Nathan P. Lawrence},
}

@inproceedings{usman2014effort,
  title={{Effort Estimation in Agile Software Development: A Systematic Literature Review}},
  author={Usman, Muhammad and Mendes, Emilia and Weidt, Francila and Britto, Ricardo},
  booktitle={International Conference on Predictive Models in Software Engineering},
  pages={82--91},
  year={2014}
}

@article{jones2004software,
  title={{Software Project Management Practices: Failure Versus Success}},
  author={Jones, Capers},
  journal={CrossTalk: The Journal of Defense Software Engineering},
  volume={17},
  number={10},
  pages={5--9},
  year={2004}
}

@article{wambeke2011causes,
  title={{Causes of Variation in Construction Project Task Starting Times and Duration}},
  author={Wambeke, Brad W and Hsiang, Simon M and Liu, Min},
  journal={Journal of Construction Engineering and Management},
  volume={137},
  number={9},
  pages={663--677},
  year={2011},
  publisher={American Society of Civil Engineers}
}

@incollection{wang2024jwst,
  title={{JWST (James Webb Space Telescope): The Most Expensive Space Telescope in History}},
  author={Wang, Jie},
  booktitle={Eye Beyond the Sky},
  pages={373--392},
  year={2024},
  publisher={Springer}
}

@techreport{jwstfinalreport2010,
    author = {Ballhaus, William F. and Casani, John and Dofman, Steven and Gallagher, David and Illingworth, Garth and Klineberg, John and Schurr, David and Lewis, Rosalind and Lobbia, Marcus},
    title = {{NASA: James Webb Space Telescope Independent Comprehensive Review Panel Final Report}},
    institution = {National Aeronautics and Space Administration},
    year = {2010}
}

@article{reichhardt2006us,
  title={{US Astronomy: Is the Next Big Thing Too Big?}},
  author={Reichhardt, Tony},
  journal={Nature},
  volume={440},
  number={7081},
  year={2006}
}

@article{shumsky1998optimal,
  title={{Optimal Updating of Forecasts for the Timing of Future Events}},
  author={Shumsky, Robert A},
  journal={Management Science},
  volume={44},
  number={3},
  pages={321--335},
  year={1998}
}

@article{kim2025information,
  title={{Information Sharing to Optimize the Wait-Time Experience}},
  author={Kim, Jeunghyun and Debo, Laurens and Shumsky, Robert A},
  journal={Social Sciences Research Network},
  year={2025}
}

@article{nafkha2016critical,
  title={{The Critical Path Method in Estimating Project Duration}},
  author={Nafkha, Rafik and Wili{\'n}ski, Artur},
  journal={Information Systems in Management},
  volume={5},
  number={1},
  pages={78--87},
  year={2016}
}

@article{thiele2025improving,
  title={{Improving Project Forecasting Accuracy by Developing the Normalised Project Management Baseline}},
  author={Thiele, Brett and Abbasi, Alireza and Ryan, Michael J},
  journal={KSCE Journal of Civil Engineering},
  volume={29},
  number={4},
  year={2025}
}

@article{jorgensen2022,
    title = {{When Should We (Not) Use the Mean Magnitude of Relative Error (MMRE) as an Error Measure in Software Development Effort Estimation?}},
    journal = {Information and Software Technology},
    volume = {143},
    pages = {106784},
    year = {2022},
    doi = {https://doi.org/10.1016/j.infsof.2021.106784},
    url = {https://www.sciencedirect.com/science/article/pii/S0950584921002263},
    author = {Magne Jørgensen and Torleif Halkjelsvik and Knut Liestøl}
}

\end{document}